\documentstyle[prl,aps,multicol,epsf]{revtex} 
\begin{document}
\title{Magnetic Phases of Electron-Doped Manganites}
\author{G. Venketeswara Pai\cite{appu} }
\address{Centre for Condensed Matter Theory, Department of Physics,\\
 Indian Institute of Science,
Bangalore - 560012, India}
\maketitle
\date{\today}
\begin{abstract}
We study the anisotropic magnetic structures exhibited by electron-doped 
manganites using a model which incorporates the  double-exchange between orbitally
degenerate $e_{g}$ electrons and the super-exchange between $t_{2g}$ electrons
with realistic values of  the Hund's coupling($J_H$),   the super-exchange
coupling($J_{AF}$), and  the bandwidth($W$).
We look at  the relative stabilities of the G, C and A type antiferromagnetic phases.
In particular we find that the G-phase is stable 
 for low electron doping as seen in experiments.
We find good agreement with the experimentally observed 
magnetic phase diagrams of electron-doped manganites
 ($x > 0.5$) such as Nd$_{1-x}$Sr$_{x}$MnO$_{3}$, Pr$_{1-x}$Sr$_{x}$MnO$_{3}$,
 and Sm$_{1-x}$Ca$_{x}$MnO$_{3}$.
We can  also explain the experimentally observed  orbital structures of  the C and A phases.
 We also extend our
calculation for electron-doped bilayer manganites of the form
R$_{2-2x}$A$_{1+2x}$Mn$_2$O$_7$ and predict that the C-phase will be absent in these systems
due to their reduced dimensionality.
\end{abstract}
\vspace{0.2cm}
\centerline{PACS : 75.30.Vn, 75.30.Gw, 75.30.Et, 71.27.+a}
\vspace{0.2cm}
\begin{multicols}{2}

\section{Introduction}
Recently there has been  a great upsurge of interest in 
doped manganites
exhibiting colossal magnetoresistance.
Most of the studies on manganites ( $R_{1-x}A_{x}MnO_{3}$ ; {\it R} = La, Nd, Pr or Sm
and {\it A} = Sr, Ca, Ba or Pb) focus on the transition from a ferromagnetic metal to
a paramagnetic insulator in the doping regime $0.15 < x < 0.4$\cite{art,coey,tvr}
which can be considered as doping the quarter filled $e_g$ band of RMnO$_3$ with holes.
More recently, {\it electron-doped} manganites\cite{eldoped}, namely systems wih $0.5 < x < 1$, 
have begun
to be explored. These seem to be different and quite interesting in their own way
with a variety of anisotropic magnetic phases and with no evidence of {\it particle
- hole symmetry}. Such systems are experimentally seen to exhibit A-type, C-type and
G-type antiferromagnetism\cite{akimoto,kajimoto,wollan},  
spectacular transitions between these phases under an applied 
magnetic field\cite{mahi} as well as possible phase separation. 
In contrast with the hole-doped systems,
there have been very few attempts 
to understand the electron-doped systems theoretically. 
 We show here that the rich magnetic phase diagram 
as well as their orbital structure
can be understood in terms
of a microscopic model which takes into account the {\it large but finite}  Hund's  rule
{\it double-exchange}(DE) coupling, effects of orbital degeneracy and the {\it super-exchange}
(SE) coupling between $t_{2g}$ spins within a band picture.\\

We study this model in detail for realistic values of the Hund's coupling, the super-exchange 
coupling and the bandwidth.
We present the phase diagram as a function of doping 
in reduced units of $J_H/t$ and $J_{AF}/t$.
From the phase diagram we deduce that the key interaction responsible for the 
stability of  
the G-phase near $x = 1.0$ 
is the super-exchange
interaction. 
We also find that the A-phase near $x \sim 0.5$ is very sensitive to the 
variation of the super-exchange interaction.
We obtain a G-type phase  for $0.85<x {\leq 1}$, a C-type phase for $0.6<x {\leq 0.85}$ 
and an A-type phase for
$0.5 {\leq}x {\leq}0.6$ for values of $J_H$ (Hund's coupling), $J_{AF}$ 
( super-exchange coupling )
and {\it t} (hopping parameter) that are in agreement with density functional calculations
\cite{satpathy}.
Thus we find that a finite value of $J_H$ leads to a magnetic phase diagram
in good agreement with experiments for a number of doped manganites $R_{1-x}A_xMnO_3$
for  $x > 0.5$.
The model also throws light on the nature of
orbital occupation of the electronic degrees of freedom 
which will lead to the experimentally observed orbital ordering\cite{kajimoto}.
We extend the mean field theory to incorporate the CE-phase at $x=0.5$ and find that it is
stabilized over a wide range of values of $J_HS_0/t$ and $J_{AF}S_0^2/t$.
We also use our model to make a number of predictions regarding the magnetic phases of
electron-doped bilayer systems R$_{2-2x}$A$_{1+2x}$Mn$_2$O$_7$ \cite{moritomo}.
Specifically, we point out that the C-phase will be absent in the electron doped
bilayer manganites due to reduced dimensionality.\\
 
We start by briefly describing the experimental situation.
In Section II we present our model. In Sections III  we present our results on the
magnetic phase diagram of the manganites. 
Section IV deals with the nature of the G-phase for low electron doping and
of canting of spins.
In Section V we explain the orbital structures observed in the C and A phases.
Section VI deals with the phase diagram of the electron-doped bilayer manganites.
Finally we make a comparison of our results with the earlier works and point out the 
shortcomings of various approaches including ours.\\

The conventional single-band double exchange model predicts a phase diagram
symmetric about $x = 0.5$. However, the behaviour of the observed ground
 state magnetic properties does not agree this simple picture.
Experiments show a remarkable asymmetry with regard to the magnetic properties of the system.
In $La_{1-x}Sr_{x}MnO_{3}$ an A - type antiferromagnetic ground state
is seen for $ 0.52 < x < 0.58$, above which it becomes a C-type antiferromagnet. 
In $Nd_{1-x}Sr_{x}MnO_{3}$
an A-type antiferromagnetic  phase extends from $ x \sim 0.5$ to $ x \sim 0.62$ and a C-type phase is
seen till $ x \sim 0.8$ \cite{kajimoto}. 
In $Pr_{1-x}Sr_{x}MnO_{3}$ the  A-type antiferromagnetism  is seen 
from $x \sim 0.48$ upto $x \sim 0.6$ 
and the  C-antiferromagnetism upto $x \sim 0.9$ \cite{akimoto}. 
The end compound $AMnO_{3}$ is a G-type antiferromagnet
and this state extends, in general, upto $x \sim 0.90$\cite{laca}. 
In particular, recent experiments on Sm$_{1-x}$Ca$_x$MnO$_3$ suggests that the G-phase,
albeit with ferromagnetic clusters embedded in them, survives upto a doping concentration
of $x=0.88$ \cite{mahi1}. Though a picture based on band-structure will not be 
appropriate in such a case, we believe that the nature of the background 
magnetic phase can still be captured since the dominant energy here is the
antiferromagnetic energy resulting from the super-exchange.

\section{Model}

If one starts from  $AMnO_{3}$ and increase the doping, the
doped electrons go into empty $e_{g}$ levels doubly degenerate in the absence of
Jahn-Teller splitting.
As noted in \cite{maezono,khomskii}  
the double-exchange between these degenerate $e_g$ levels 
along with the
super-exchange between $t_{2g}$ core spins lead to a qualitatively different phase diagram 
which is highly asymmetric about $x=0.5$. 
 However, the resulting $T=0$ phase diagram they obtained while asymmetric is in 
disagreement with experiments on several counts (see below).
This motivates a detailed  study of an orbitally degenerate
double-exchange(DE) and super-exchange(SE) model for $0.5 < x < 1$ with  realistic values of
parameters $J_H, J_{AF}$ and bandwidth $W$ \cite{jh}.
The effective Hamiltonian describing the low energy properties of the system is
\begin{eqnarray}
H&=&J_{AF} \sum_{<ij>}{\bf S}_i \cdot {\bf S}_j
 - J_H \sum_{i,\alpha,\mu,\mu{^\prime}} {\bf S}_i \cdot c_{i \alpha \mu
}^{\dag}
     \vec{\sigma}_{\mu \mu{^\prime}} c_{i \alpha \mu{^\prime}}
 \nonumber\\
 &-&\sum_{<ij>,\mu} t_{ij}^{\alpha \beta}
                       c_{i \alpha \mu}^{\dag} c_{j \beta \mu}
\label{eq:H}
\end{eqnarray}
Here $\alpha$ and $\beta$ denote the $d_{3z^{2}-r^{2}}$ and $d_{x^{2}-y^{2}}$ orbitals respectively,
$S_{i}$ is the $t_{2g}$ spin at site {\it i}, $J_{H}$ is the Hund's coupling and $J_{AF}$ the 
super-exchange
between $t_{2g}$ spins at nearest - neighbour sites {\it i} and {\it j}, and $\mu$ stands for the
spin degree of freedom of the itinerant electrons.
The hopping matrix elements are determined by the symmetry of $e_g$
orbitals\cite{phil}.\\

We treat the spin subsystem quasiclassically. Assuming a homogeneous
ground state
we take $ {\bf S}_{i} = {\bf S}_0\hspace{2mm}  \cos(\vec Q. \vec r_i)$.
where
${\vec Q} = (0, 0, 0)$ for the ferromagnetic phase,
${\vec Q} = (\pi, \pi, \pi)$ for the G-type antiferromagnetic phase,
${\vec Q} = (\pi, \pi, 0)$ for the C-type antiferromagnetic phase, and
${\vec Q} = (0, 0, \pi)$ for the A-type antiferromagnetic phase.   
Canting can be included by assuming ${\bf S}_{i} = S_{0}
(\sin\theta_{i}, \sin\theta_{i}, \cos\theta_{i})$ with
$\theta_{i}$ taking values between 0 and $\pi$. This is discussed late below \cite{cant}.\\

Under these assumptions the electronic part of the Hamiltonian reduces to 
\begin{eqnarray}
H_{el}&=& \sum_{k, \alpha, \beta} \epsilon_{k}^{\alpha \beta}c_{k \alpha \uparrow}^{\dag}c_{k \beta \uparrow} + 
\sum_{k, \alpha, \beta} \epsilon_{k}^{\alpha \beta}c_{k \alpha \downarrow}^{\dag}c_{k \beta
\downarrow}  \nonumber\\
&-& {J_{H}S_{0} \over 2} \sum_{k, \alpha}c_{k \alpha \uparrow}^{\dag}c_{k+Q \alpha \uparrow} 
   - {J_{H}S_{0} \over 2} \sum_{k, \alpha}c_{k \alpha \uparrow}^{\dag}c_{k-Q \alpha \uparrow} 
\nonumber\\
&+& {J_{H}S_{0} \over 2} \sum_{k, \alpha}c_{k \alpha \downarrow}^{\dag}c_{k+Q \alpha \downarrow}
   + {J_{H}S_{0} \over 2} \sum_{k, \alpha}c_{k \alpha \downarrow}^{\dag}c_{k-Q \alpha \downarrow}
\end{eqnarray}
with \cite{phil}
\begin{eqnarray}
\epsilon_{11}&=& -\frac{2}{3} t(\cos k_x + \cos k_y)
         -\frac{8}{3} t \cos k_z, \nonumber \\
\epsilon_{12}&=& \epsilon_{21}= -\frac{2}{\sqrt{3}} t(\cos k_x - \cos k_y) \nonumber \\
\epsilon_{22}&=& -2t(\cos k_x + \cos k_y).   
\label{eq:hop}
\end{eqnarray}
The super-exchange contribution to the Hamiltonian is given by 
\begin{eqnarray}
H_{SE}&=& {J_{AF}S_{0}^{2} \over 2} ( 2\cos \theta_{xy} + \cos \theta_z)
\end{eqnarray}
with $\theta_{xy}$ = $\theta_{z}$ = 0 for ferromagnetic,
      $\theta_{xy}$ = $\theta_{z}$ = $\pi$ for the G -type  antiferromagnetic,
	$\theta_{xy}$ = $\pi$ and $\theta_{z}$ = 0 for the C -type  antiferromagnetic, and
$\theta_{xy}$ = 0 and $\theta_{z}$ = $\pi$ for the A -type  antiferromagnetic phases.
 Here $\theta_{xy}$ is the angle between nearest neighbour spins in the {\it x-y} plane, and $\theta_{z}$
is the angle between nearest neighbour spins in the {\it z} direction.
Inclusion of canting by assuming $ {\bf S}_{i} = S_{0}
(\sin\theta_{i}, \sin\theta_{i}, \cos\theta_{i})$ will
connect different spin species at the same site. These contibutions come from the
$\sigma_{x}$ and $\sigma_{y}$ terms in the DE part of the Hamiltonian which are absent
when canting is absent. Thus the DE part of the Hamiltonian becomes
\begin{eqnarray}
H_{DE}&=& -J_{H}S_{0} \sum_{j, \alpha} \cos\theta_{j} (c_{j \alpha \uparrow}^{\dag}c_{j \alpha \uparrow} - c_{j \alpha \downarrow}^{\dag}c_{j \alpha \downarrow}) \nonumber\\
&-& J_{H}S_{0} \sum_{j, \alpha} \sin\theta_{j} (c_{j \alpha \uparrow}^{\dag}c_{j \alpha \downarrow} + c_{j \alpha \downarrow}^{\dag}c_{j \alpha \uparrow}) \nonumber\\
&+& J_{H}S_{0} \sum_{j, \alpha} i \sin\theta_{j} (c_{j \alpha \uparrow}^{\dag}c_{j \alpha \downarrow} - c_{j \alpha \downarrow}^{\dag}c_{j \alpha \uparrow})
\end{eqnarray}

We have neglected the correlation term in the Hamiltonian 
$U\sum_{i \alpha}n_{i, \alpha \uparrow}n_{i, \alpha \downarrow} + 
 V\sum_{i, \sigma, \nu}n_{i, 1, \nu}n_{i, 2, \sigma}$
and the Jahn - Teller contribution 
$g\sum_{i,  \alpha, \beta, \sigma}c_{i, \sigma, \alpha}^{\dag} 
{\bf Q}_{i}^{\alpha \beta}
c_{i, \sigma, \beta}$ with ${\bf Q}$ describing the local distortion which lifts the degeneracy
\cite{millis}.
We neglect the correlation term because of the low electron doping regime we are interested in.
($x = 0.5$ refers to a filling of 0.125 in our model and the filling ranges from $0$ to
$0.125$). For the same reason the inter-site Coulomb correlations, which may be necessary for
the stability of charge ordered phase around $x = 0.5$, are also neglected.
A cooperative Jahn-Teller effect can drastically change the magnetic ground state\cite{feinberg}.
However,
since the carrier concentration is very small, so is the {\it effective number of Jahn - Teller
centres} and 
hence we do not expect any 
qualitative change in the magnetic phase diagram though both may be required 
alongwith the breathing mode distortions induced by holes to explain 
the CE - type charge ordered phase at $x = 0.5$ \cite{venkat}. 
Doping-induced disorder can have two effects. Firstly substitutional
disorder may localize $e_g$ electrons. However as long as the localization length
is more than the inter atomic spacing, the hopping to nearest neighbour sites
will split the energy levels into bonding and antibonding orbitals with electrons
occupying the bonding orbitals. This process is naturally taken care of in our
model. 
Secondly the presence of a magnetic rare-earth ion can have coupling with the
magnetic $Mn^{3+}$ ion and thus leading to change in the Mn-RE coupling as doping
varies. However in most of the manganites, the RE ion in general is non-magnetic ({\it eg. La}) 
except, say in Pr. However studies on Pr-Sr system around $ x = 0.37 $
\cite{hwang} have shown that Mn-Pr coupling plays no role in the magnetic 
propertis.
Hence we also do not expect substitutional disorder to play a role in determining
the magnetic phases though, as argued in \cite{maezono,khomskii}, it might play
a role in the transport properties of these compounds.
We  obtain
the magnetic phase diagram by minimizing the total energy
$H_{el}+H_{SA}$ as a function of filling by fixing the chemical potential.\\

We present the magnetic phase diagram for both manganites and bilayer manganites
as a function of $J_H/t$ and $J_{AF}/t$.
From density functional studies \cite{satpathy}
we estimate $t = 0.15$eV, $J_{H}S = 0.75$eV and
$J_{AF}S^{2} = 8$meV. Hence, we choose ${J_{H}S_{0}/t}$ = 5 and ${J_{AF}S_{0}^{2}/t}$ =
0.053.
This value of $J_{AF}S_{0}^2/t$ also leads to the correct  mean field $T_N$ for the
end compound CaMnO$_3$. Hence we use the phase diagram corresponding to these values
for making comparison with experiments\cite{rahul}.            
\section{Phase Diagram of the Electron Doped Manganites}
The $x=1$ limit corresponds to empty $e_g$ orbitals. The only contribution to the
Hamiltonian comes from the SE interaction which is isotropic and hence  leads to the
G-phase at $x=1$. At low electron doping, however, the SE still wins over the Hund's coupling
and leads to the G-phase.

Doped electrons go into states with minimal energy
corresponding to the $\Gamma$ point at ${\bf k} =0$.
(This is a consequence of the band-picture we, as well as
other workers, use. The possibility of the doped
electron forming ferromagnetic clusters is mentioned later, in which case the
band-picture will break down.)
We first assume that all doped charges go into the state with ${\bf k} = 0$
and neglect for the moment the effects due to finite filling of the bands.
(Strictly speaking this is only the case for very small doping).
Assuming uncanted states  the energies  for various magnetic states
at ${\bf k} = 0$ are
\begin{eqnarray}
E_G&=& -3J_{AF}S_{0}^2/2t-y \sqrt{16+(J_{H}S_{0}/2t)^2}, \\
E_A&=& J_{AF}S_{0}^2/2t-4y-J_{H}S_{0}y/2t, \\
E_C&=& -J_{AF}S_{0}^2/2t-8y/3-y \sqrt{16/9+(J_{H}S_{0}/2t)^2}, \\
E_F&=& 3J_{AF}S_{0}^2/2t -4y - J_{H}S_{0}y/2t.
\label{eq:energystate}
\end{eqnarray}
Here $y$ is the actual electron filling in the two band model and is related to
$x$ as
$x=-4y+1$.                         
For $J_{H}S_{0}/t = 5$ and $J_{AF}S_{0}^{2}/t = 0.053$ we find that
the  G-phase is stable upto $x=0.76$ 
 beyond which
the A-phase becomes stable. In Fig(1) we present the phase diagram assuming the electrons go into
the $\Gamma$ point and there is no canting of core spins. We plot the phase diagram as a
function of doping and $J_{AF}S_0^2/t$ for a fixed value of $J_HS_0/t =5$ and doping and $J_HS_0/t$
for a fixed value of $J_{AF}S_0^2/t = 0.053$.\\

 The effects due to finite band-filling will alter these
values and the numerically obtained values can be read from Fig(2) and Fig(3). 
This also leads to the physically expected result that the doping region
over which the G-phase stabilizes grows with $J_{AF}/t$.
As electron doping increases the kinetic energy starts dominating over the SE contribution leading to increased spin alignment.
This happens because kinetic energy is an increasing function of doping
and for small doping it is proportional to the electron filling.
However, a three-dimensional antiferromagnetic spin alignment does not allow 
for the motion of electrons. So to take advantage of the kinetic energy gain
the mobile electrons polarize the spins along chains, planes and finally in all three directions successively.
The reduction in the DE energy due to such alignment is overcome by the gain
in kinetic energy beyond some doping value (for given values of $J_H/t$ and
$J_{AF}/t$) and this point defines the G-C phase boundary. Moreover, as we will see in the next section, the C-phase has orbital ordering of $d_{z^2}$-type
and the A-phase has orbital ordering of $d_{x^2-y^2}$-type. Thus the interplay
of the spin alignment along chains or planes and the corresponding orbital
order also leads to change of the `{\it effective hopping parameters}', $t_z$
and $t_{xy}$, in the $z$ and $x-y$ directions.
In general, this leads to  the system transforming from 
one-dimensional, to two-dimensional and finally three dimensional ferromagnetic structures
with increasing doping.
Thus the competition between   
{\it effective kinetic energy} (determined by $J_{H}$ and band filling) and 
{\it super-exchange} leads to transitions G-C-A-F(with number of
antiferromagnetic bonds  6, 4, 2 and  0 respectively)
 as the doping is varied
for a given $J_H/t$.\\

 In Fig(2) we present the results for ${J_{H}S_{0}/t}$ = 5. 
For ${J_{AF}S_{0}^{2}/t}$ = 0.053 we find that 
the system has a stable ferromagnetic ground state upto
$x = 0.47$,
the A-phase is favoured for $ x < 0.57$,
the C-phase upto 0.85. The G-phase becomes the stable phase for $0.85 < x < 1$.
We also find that the A - phase near $x = 0.5$ is stable only for a limited range of
${J_{AF}S_{0}^{2}/t}$. 
The overall phase diagram is in excellent agreement with the experimentally
observed phase diagram of NdSr, PrSr and SmCa systems.

In Fig(3) we present the results for ${J_{AF}S_{0}^{2}/t}$ 
= 0.053. We find that the A-phase, stable near $x = 0.5$ for smaller values of
${J_{H}S_{0}/t}$, gets pushed to the right making the ferromagnetic state stable near 
$x = 0.5$ for large values of $J_H/t$. 
However, in contrast to the earlier case, the A - phase is stable over a wide
range of values of ${J_{H}S_{0}/t}$. We conclude that  the A - phase near $x = 0.5$ is
very sensitive to the variation of ${J_{AF}S_{0}^{2}/t}$ and rather less sensitive to 
the variation of ${J_{H}S_{0}/t}$.\\

At $x=0.5$ most of the manganites have a charge/orbital ordered ground state with the
magnetic phase being the CE-type antiferromagnet. Our mean field theory can be
extended to include the uncanted CE phase by assuming 
${\bf S}_i = {\bf S}_0/2 \hspace{0.2cm}(\cos(\vec Q_1. \vec r_i)-\cos(\vec Q_2. \vec r_i)+
\cos(\vec Q_3. \vec r_i) + \cos(\vec Q_4. \vec r_i))$ with
${\vec Q_1} = (\pi, 0, \pi)$, ${\vec Q_2} = (0, \pi, \pi)$, ${\vec Q_3} = (\pi/2, 3\pi/2, \pi)$
and ${\vec Q_4} = (3\pi/2, \pi/2, \pi)$.
We present the phase diagram at $x=0.5$ as a function of $J_HS_0/t$ and $J_{AF}S_0^2/t$
in Fig(4). We find that at $J_HS_0/t = 5$ and $J_{AF}S_0^2/t = 0.053$, the CE phase 
stabilizes over other phases. In fact, the CE phase is stabilized over a wide region
of the phase diagram at $x=0.5$. This may explain why most of the manganites at $x=0.5$ have 
the CE phase as the magnetic ground state. However, it is to be noted that the CE phase 
we obtain is not charge/orbital ordered. Other interaction such as the strong Coulomb 
repulsion or coupling of the lattice degrees of freedom to the $e_g$ electrons may be needed
to make this phase charge/orbital ordered. 
There are contrasting views regarding the origin of the charge/orbital ordered
CE phase and the precise role of JT and Coulomb effects is still not clear. 
Strong on-site Coulomb correlations within a two-band model seem to stabilize the CO phase
at $x=0.5$ \cite{khomskii1}.
It can also be thought of as emerging due to the doping dependent
Berry phase associated with the JT effect \cite{takada}. 
However, manganites at $x=0.5$ exhibit a variety of ground states including the CE phase as
in PrCa or NdSr, the A-phase as in PrSr or the metallic ferromagnetism as in LaSr. 
A Monte-Carlo study of the two-band model with JT phonons \cite{dagottoco} 
seem to capture most of these phases. 
An extension of our mean field theory incorporating the Jahn-Teller effect and breathing mode
reproduces the charge/orbital ordered CE phase as well as the A-type phase \cite{venkat}.\\

\section{The nature of the G-phase and Canting}
Expermentally \cite{akimoto}, \cite{kajimoto} it is seen that there seems to be little canting
in the A and C phases. This was also emphasized by Maezono {\it et. al.} \cite{maezono}.
It is
also seen that there is a predominant occupation of orbitals of one character
in these phases.
Recent experiments by Mahendiran {\it et. al.}\cite{mahi1}
on Sm$_{1-x}$Ca$_x$MnO$_3$
suggest that even the G-phase for low doping may have little canting.
The doped carriers seem to form ferromagnetic clusters leaving behind a
uniform G-phase as background.                                       
In the band-picture, we have already noticed that for low electron doping the SE wins over the
DE and the phase is G-type antiferromagnetic.
One expects this phase
to be canted as electrons gain kinetic energy
due to the DE mechanism. The canting angle will be anisotropic, i.e., $\theta_{xy}$ will be
different from $\theta_z$ due to the anisotropy of the
hopping integrals $t_{\alpha \beta}^{ij}$. However, no specific orbital ordering can be seen in 
this phase.This phase (without any orbital ordering) also has to be contrasted to
the A-phase near $x = 0.5$ which has orbital ordering of $d_{x^2-y^2}$ type (see next section). 
The stability of the G-phase near $x = 1$ is because of the dominance of
antiferromagnetic energy whereas the stability of the A-phase near $x = 0.5$ arises from
the kinetic energy gain through DE in the plane due to selective
$d_{x^2-y^2}$ orbital ordering.    
Moreover, for finite $J_H$ the canting is relatively small leading to a phase which closely
resembles the G-phase at $x=1$. 
In Fig(5) we plot the canting angles as a function of $J_H/t$ for a fixed value
of $J_{AF}/t$ for some representative value of doping($x = 0.98$). We find that the canting
angle increases as a function of $J_H/t$ for a given filling and $J_{AF}/t$
near $x=1$.\\                        

In the limit $J_H \rightarrow \infty$, electron hopping to neighbouring sites
with antiparallel core spins is not allowed. This is because the `effective
hopping parameter' for $J_H \rightarrow \infty$ is proportional to
$tcos(\theta/2)$ where $\theta$ is the angle between the spins at neighbouring
sites and antiparallel arrangement of spins reduces the `effective hopping parameter' to zero.
Hence the only way the electrons can take advantage of the kinetic energy gain due to increased doping
is by canting the spins as much as possible. However, since $t_{ij}$'s 
are anisotropic the canting angles will also be anisotropic. For a representative value
of electron doping ($x = 0.98$) we find that  there is  no canting
in the $z$ direction  
and spins cant by about $10^o$ in the $x-y$ plane.
This gives rise to a net ferromagnetic moment
in the plane with a value higher than that across the layers. Hence one would
think of it as a canted A-phase as in \cite{khomskii}. However, inclusion
of finite $J_H$ changes this picture. A finite value of $J_H$ allows the spins to go
to `{\it wrong spin state}' at neighbouring site with an energy cost $J_H$. 
Hence the canting angle is reduced drastically compared to the $J_H \rightarrow
\infty$ limit. In fact, for experimentally realistic values of $J_H$ the canting is
almost absent for low electron doping as can be inferred from Fig(5).
(In fact one expects no canting for $J_H = 0$ as DE is not operative.)
Moreover, the kinetic energy gain which is proportional 
to the doping is also not effective in overcoming the SE energy. Hence one gets
a canted G-phase with very small canting angles, thus resembling the G-phase at $x = 1$. 
Since the kinetic energy gain is also very small due to the smallness of the canting angle, this phase does not have any preferential orbital arrangement
of the $d_{z^2}$ or $d_{x^2-y^2}$ type as in the C and A phases. Thus we find that
the stability of the G-phase is mainly due to the dominance of SE energy. This
also means that the doping region over which the G-phase stabilizes will grow
with increase in $J_{AF}/t$. In particular, for $J_{AF}/t = 0$ the system
should exhibit ferromagnetism for any doping making the G-C phase boundary collapse to
the $x=1$ point in the $J_{AF}/t$-$x$ phase plane.
However \cite{khomskii} find that the phase boundary between the canted G-phase
and the C-phase does not change significantly as $J_{AF}/t$ is varied. 
More surprisingly, their phase diagram, if extrapolated to $J_{AF}/t =0 $,
will give the canted A-phase over a small region of doping near $x = 1$.
In contrast to this, our phase diagram  gives a ferromagnetic state for
$J_{AF}/t = 0$ for the whole doping regime and the stability region of the
G-phase grows with increase in $J_{AF}/t$ in agreement with the
physically expected result.
Our results  agree in general with
the results of Maezono {\it et. al} \cite{maezono}
though the A-phase near $x \sim 0.5$ is missing in that
work.  Sheng and Ting \cite{sheng} considered the problem from the strong correlation limit
in contrast to the band limit which we have adopted. The C-phase between $x=0.6$ and
$x=0.9$ is missing in the strong correlation limit.\\

\section{Orbital Structure}

We find that in the C-phase the occupied orbitals are predominantly of
$d_{z^2}$ character with a small admixture of $d_{x^2-y^2}$.
This happens because the electrons
gain kinetic energy along the direction in which ferromagnetic correlations
are stronger. For the same reason we find that in the A-phase
the occupied orbitals are predominantly of
$d_{x^2-y^2}$ character.
This, in effect, leads to suppression of hopping along antiferromagnetic bonds and
 explains
why there is  little canting  in these systems.
 This is in agreement with experiments on Nd$_{1-x}$Sr$_{x}$MnO$_3$
\cite{kajimoto}.
 This also leads to a highly anisotropic band structure
for G, C and A type structures and this feature becomes sharper as $J_H$ increases.
In particular, the C-phase  has a quasi-one dimensional density of states.
This also makes this phase very sensitive to substitutional disorder, possibly
making it insulating.                 
However, the A-phase is not sensitive to  disorder and this 
rationalizes the (in-plane) metallic A-phase seen in experiments\cite{akimoto}.
The nature of the occupied orbitals prevents electron motion along the $z$ direction
giving rise to a large anisotropy in the in-plane and out-of-plane resistivities.
 Experiments which
probe the density of states, like tunneling measurements, will be able to see this feature. 
The low-temperature magnon spectrum will also throw light on the precise nature
of the antiferromagnetic phase near $x = 1$ and specifically the nature of canting 
in different manganites.\\ 

\section {Phase Diagram of the Electron Doped Bilayer  Manganites}

The present scheme of calculation can also be applied to electron - doped bilayer
manganites such as $R_{2-2x}A_{1+2x}Mn_{2}O_{7}$ about which very little is known\cite{moritomo}.
Since the 
interlayer coupling is roughly two orders of magnitude smaller than the coupling between 
bilayers one can apply the degenerate  double - exchange, super - exchange model for a
two layer system to study bilayered manganites.
 In this case the Brillouin  zone is modified with $k_z$ taking only two values.
As noted earlier the magnetic structure depends on the competition between
the super-exchange and the kinetic energy renormalized by magnetic structure and orbital
degrees of freedom. This suggests that in bilayer compounds where the kinetic energy gain is 
predominantly in planes than in the {\it z} direction, the A-type antiferromagnetic phase is
stabilized over the C-phase.
This means that the dimensionality of the system plays a crucial
role in the stability of  the C-phase.
This can be clearly seen in the limit $J_H \rightarrow \infty$ where the band structure for
C-phase becomes one dimensional with
$\epsilon = -{8 \over 3}t \cos(k_z)$. Detailed calculations support this picture as seen
in Fig(6) where we present the results for a fixed value of $J_{H}S/t = 5$ and
in Fig(7) where we present the results for a  fixed value of $J_{AF}S^{2}/t = 0.053$.
Battle {\it et. al.} \cite{battle} have reported an A-type phase for NdSr$_2$Mn$_2$O$_7$ ($x = 0.5$) and
Nd$_{1.1}$Sr$_{1.9}$Mn$_2$O$_7$ ($x=0.45$). We believe that this phase should extend even beyond $x=0.5$  in accordance with our picture. Our phase diagram is in accordance with that of Maezono and Nagaosa 
\cite{nagaosa}

\section{Discussion}
It is interesting to study the phase transitions between these anisotropic structures under an 
applied magnetic field in $z$ direction. We find that the   G-type phase becomes a canted A-type phase
before transforming to the  ferromagnetic phase for large $x$ (close to 1).
This is in agreement with recent experiments \cite{mahi}. Further study is needed in this
direction covering the whole doping regime $0.5 < x < 1$. 

A major drawback  of the current approach as well as that of earlier works
is the homogeneous magnetic phases they predict. It seems
likely that a phase separated regime is energetically more favourable than the canted phase
\cite{dagsc}. 
Phase separation, static or dynamic, seems to be a notable feature of manganites 
in the low-hole doped regime, charge ordered regime as well as the intermediate regime
where there is a ferromagnetic metal to paramagnetic insulator transition as the temperature
is varied. 
 Batista {\it et. al.} \cite{batista} through exact diagonalization studies of a
single band model on small one dimensional clusters find that non-uniform ground states
are highly possible in DE-SE systems. In particular, they find that at low electron
doping, doped carriers get trapped at impurity sites and form ferromagnetic clusters.
It will be interesting to study the two-band model to find exact nature of the
G-phase near $x=1$. We expect the ferromagnetic clusters to be anisotropic
  in size with `$x-y$ radius' being larger than the `$z$ radius'.
It should be possible to study phase separation using an orbitally degenerate version of
the continuum model proposed by Soto et.al. \cite{soto}.  It is also possible that
 the spiral \cite{spiral}
and the flux phases  \cite{flux}
get stabilized for some values of doping
as in the case of a single band double-exchange model though in our mean field picture
we have not considered these phases.
Work along these lines is in
progress and will be reported elsewhere.\\

To compare our results with the earlier work, we find a  G-phase for
low electron doping.
We also find that the region over which the G-phase is stabilized increases with
$J_{AF}/t$. This feature survives when canting is included as the canting angle is small for 
finite $J_{H}/t$.
Our mean field theory takes into account the canting of core spins and also results in the
A-phase near $x=0.5$ (as seen in experiments) both of which are missing in the work
of Maezono {\it et. al.} \cite{maezono}. Our model concentrates on the  minimum
number of relevant parameters and gives a unified 
picture of the electron-doped manganites(including 
bilayers).
This is in sharp contrast to the work of Maezono {\it et. al.} which uses five 
dimensionless parameters and separate order parameters for magnetic and orbital ordering. 
Our mean field theory also reproduces the C-phase between $x=0.6$ and $x=0.9$ which is
missing in the strong coupling limit of 
Sheng and Ting \cite{sheng}.
We also clarified the nature of the G-phase near $x=1$ and the G-C phase boundary is as
expected on physical grounds in contrast to van den Brink and Khomskii \cite{khomskii}.


In conclusion  we have studied a model for electron-doped manganites with super-exchange 
between $t_{2g}$ electrons and double-exchange between orbitally degenerate $e_{g}$
electrons. We find that finite $J_{H}$ changes the phase diagram qualitatively. In particular
the G-phase is favoured for low electron doping. 
This happens because the $finite-J_H$ model, by allowing electrons to hop
to neighbouring sites at an energy  cost of $J_H$ reduces the canting
making the phase resemble more to the G-phase.
The phase diagram agrees very well with
the experimental phase diagram of manganites for $0.5 < x < 1$.
By extending our mean field theory to incorporate the CE phase we find that it is stabilized 
over a wide range of values of $J_HS_0/t$ and $J_{AF}S_0^2/t$ at $x=0.5$.
We extended this model for
a two-layer system to predict  the magnetic phase diagram of electron doped bilayer manganites.
Here we find that the reduced dimensionality washes out the C-type phase. 
We also notice that the kinetic energy gain due to DE leads to selective
orbital ordering in the A and C phases while it is absent in the G-phase.
We conclude that the present model qualitatively explains the anisotropic magnetic phases
and believe that it can describe the  phase transitions between these structures under an external
field. 
A detailed study of this model is called for which should reveal
the speculation about the phase separation in electron doped manganites. 
\section{Acknowledgements}
 I thank
T.V.Ramakrishnan for discussions,  critical comments and his encouragement,
R. Mahendiran  for discussions and comments, A. Pande for a careful reading of 
the manuscript, M. Mithra for help with figures,
CSIR (India) for support and SERC (IISc, Bangalore) for 
computational resources.

Fig(1) : Phase diagram of the double-exchange and super-exchange model with
degenerate $e_g$ orbitals assuming the doped electrons go into the $\Gamma$ point and
there is no canting of the core spins. (a) Phase diagram as a function of $J_{AF}S_0^2/t$
for a fixed value of $J_HS_0/t = 5$. (b) Phase diagram as a function of $J_HS_0/t$ for a
fixed value of $J_{AF}S_0^2/t = 0.053$.\\
\vspace{0.3cm}\\
Fig(2) : Phase diagram of the double-exchange and super-exchange model with
degenerate $e_g$ orbitals for a fixed value of $J_{H}S_{0}/t = 5$.
Depending on the electron doping concentration and the ratio of the $t_{2g}$
superexchange to the $e_{g}$ bandwidth, $J_{AF}S_{0}^2/t$,
we find the A-type, C-type, G-type
or ferromagnetic order. Values of $J_{H}$, t and $J_{AF}$ were taken from
density functional  calculation[9].\\
\vspace{0.3cm}\\
Fig(3) :  Phase diagram of the double-exchange and super-exchange model with
degenerate $e_g$ orbitals for a fixed value of $J_{AF}S_{0}^2/t= 0.053$.
Depending on the electron doping concentration and the ratio of the
Hund's coupling to the  $e_{g}$ bandwidth, $J_{H}S_{0}/t$
we find the A-type, C-type, G-type
or ferromagnetic order. Values of $J_{H}$, t and $J_{AF}$ were taken from
the density functional  calculation[9].\\
\vspace{0.3cm}\\
Fig(4) : Phase diagram of the double-exchange and super-exchange model with 
degenerate $e_g$ orbitals at $x=0.5$.\\

\vspace{0.3cm}
Fig(5) : The angle difference between the neighbouring spins (in radians)
for a representative value of doping $x=0.98$ and $J_{AF}/t = 0.053$ as a
function of $J_{H}/t$.\\ 
\vspace{0.3cm}\\
Fig(6) : Phase diagram of the bilayer system for a fixed value of $J_{H}S_{0}/t
= 5$.
Depending on the electron doping concentration and the ratio of the $t_{2g}$ 
superexchange to the
$e_{g}$ bandwidth, $J_{AF}S_{0}^2/t$,
we find the A-type, G-type or ferromagnetic  order.
Note that the C-phase is missing in the bilayer system.\\
\vspace{0.3cm}\\
Fig(7) :  Phase diagram of the bilayer system for a fixed value of $J_{AF}S_{0}^
2/t = 0.053$.
Depending on the electron doping concentration and the ratio of the
Hund's coupling to the  $e_{g}$ bandwidth, $J_{H}S_{0}/t$
we find the A-type or G-type  phases.\\
\end{multicols}{2}

\begin{references}
\bibitem[\ast]{appu} {\it email} : {\small venkat@physics.iisc.ernet.in}
\bibitem{art} A. P. Ramirez, Jl. Phys. Cond. Matter {\bf 9}, 8171 (1997).
\bibitem{coey} J. M. D. Coey, M. Viret and S. von Molnar,  Adv. in Physics {\bf 48}, 167
(1999). 
\bibitem{tvr} T. V. Ramakrishnan, in {\it Colossal Magnetoresistance, Charge
Ordering and Related Properties of Manganese Oxides} ed. by C.N.R. Rao
and B. Raveau, World Scientific (1998).
\bibitem{eldoped} In this paper we use the term electron-doping for compounds with
$x > 0.5$. This should not be confused with doping $AMnO_{3}$ with a tetravalent
 ion such as Ce.
\bibitem{akimoto} T. Akimoto, Y. Maruyama, Y. Moritomo, A. Nakamura, K, Hirota, K. Ohoyama
and M. Ohashi, Phys. Rev. {\bf B}  {\bf 57},R5594 (1998).
\bibitem{kajimoto} R. Kajimoto, H. Yoshizawa, H. Kawano, H. Kuwahara, Y. Tokura, K. Ohoyama
and M. Ohashi, Phys. Rev. {\bf B} {\bf 60}, 9506 (1999).
\bibitem{wollan} E. O. Wollan and W. C. Koehler, Phys. Rev. {\bf 100}, 545 (1955).
\bibitem{mahi} R. Mahendiran, private communication.
\bibitem{satpathy} S. Satpathy, Z. S. Popovi\'{c} and F. R. Vukajlovi\'c,
Phys. Rev. Lett. {\bf 76}, 960 (1996).
\bibitem{moritomo} Y. Moritomo, A. Asamitsu, H. Kuwahara and Y. Tokura, Nature {\bf 380}, 141 (1996).
\bibitem{laca} Our results cannot be applied to systems such as La$_{1-x}$Ca$_{x}$MnO$_3$
for which the C-phase extends over a large range of doping.      
\bibitem{mahi1} R. Mahendiran, M. R. Ibarra, A. Maignan, C. Martin, B. Raveau
and C. Ritter, manuscript under preparation.
\bibitem{maezono} R. Maezono, S. Ishihara and N. Nagaosa, Phys. Rev. {\bf B} {\bf 57}, R13993 (1998
).                               
\bibitem{khomskii} J. van den Brink and D. Khomskii, Phys. Rev. Lett. {\bf 82}, 1016 (1999).
\bibitem{jh} Though the $J_H \rightarrow \infty$ limit explains several experiments well,
e.g., the low temperature magnon spectrum, finite $J_H$ effects are also important.
For example, the interband transitions observed in optical conductivity measurements
are between bands spin-split by a gap of $O(J_H)$. This happens because the energy scales
of $J_H$ and $W$ are comparable in these systems which can also be inferred from
electronic band structure calculations.
\bibitem{phil} P. W. Anderson, Phys. Rev. {\bf 115}, 2 (1959); 
	K. I. Kugel and D. Khomskii, Sov. Phys. JETP {\bf 37}, 725 (1973).
\bibitem{cant} When canting is included, we define the C-phase as fully ferromagnetic in
$z$ direction, the A-phase as fully ferromagnetic in $x-y$ plane and the G-phase as the structure
with spins canted from istropic antiferromagnet. However, [14] defines the A-phase
with $\cos \theta_{xy} > \cos \theta_{z}$ and the C-phase with 
$\cos \theta_{xy} < \cos \theta_{z}$ where $\theta_{xy}$ and $\theta_{z}$ are the canting angles.
\bibitem{millis} A. J. Millis, P. B. Littlewood and B. I. Shraiman, Phys. Rev. Lett. {\bf 75}, 5144 
(1995).
\bibitem{feinberg} D. Feinberg, P. Germain, M. Grilli and G. Seibold, Phys. Rev. {\bf B} {\bf 57},
R5583 (1998).
\bibitem{venkat} G. Venketeswara Pai, under preparation. 
\bibitem{hwang} H. Y. Hwang, P. Dai, S-W. Cheong, G. Aeppli, D. A. Tennant and
H. A. Mook, Phys. Rev. Lett {\bf 80}, 1316 (1998).        
\bibitem{rahul} The precise value of $J_H/t$ is hard to obtain. Various values have been used 
in the literature. We follow \cite{satpathy} and S. K. Mishra, R. Pandit and S. Satpathy,
Phys. Rev. {\bf B} {\bf 56}, 2316 (1997).
\bibitem{khomskii1} J. van den Brink, G. Khaliullin and D. Khomskii, Phys. Rev. Lett. 
{\bf 83}, 5118 (1999). 
\bibitem{takada} T. Hotta Y. Takada, H. Koizumi and E. Dagotto, Phys. Rev. Lett. {\bf 84}, 2477
 (2000).
\bibitem{dagottoco} S. Yunoki, T. Hotta and E. Dagotto, Phys. Rev. Lett. {\bf 84}, 3714 (2000).
\bibitem{sheng} L. Sheng and C. S. Ting, {\tt cond-mat}/9812374. 
\bibitem{battle} P. D. Battle, M. A. Green, N. S. Laskey, J. E. Millburn, P. G. Radaelli,
M. J. Rosseinsky, S. P. Sullivan and J. F. Vente, Phys. Rev. {\bf B} {\bf 54}, 15967 (1996).
\bibitem{nagaosa} R. Maezono and N. Nagaosa, Phys. Rev. {\bf B} {\bf 61}, 1825 (2000).
\bibitem{dagsc} A. Moreo, S. Yunoki and E. Dagotto, Science {\bf 283}, 2034 (1999).
\bibitem{batista} C. D. Batista, J. Eroles, M. Avignon and B. Alascio,
Phys. Rev. {\bf B} {58}, R14689 (1998).
\bibitem{soto} J. M. Rom\'{a}n and J. Soto, Phys. Rev. {\bf B} {59}, 11418 (1999).
\bibitem{spiral} J. Inoue and S. Maekawa, Phys. Rev. Lett. {\bf 74}, 3407 (1995).
\bibitem{flux} M. Yamanaka, W. Koshibae and S. Maekawa, Phys. Rev. Lett. {\bf 81}, 5604 (1998).
\end{references}
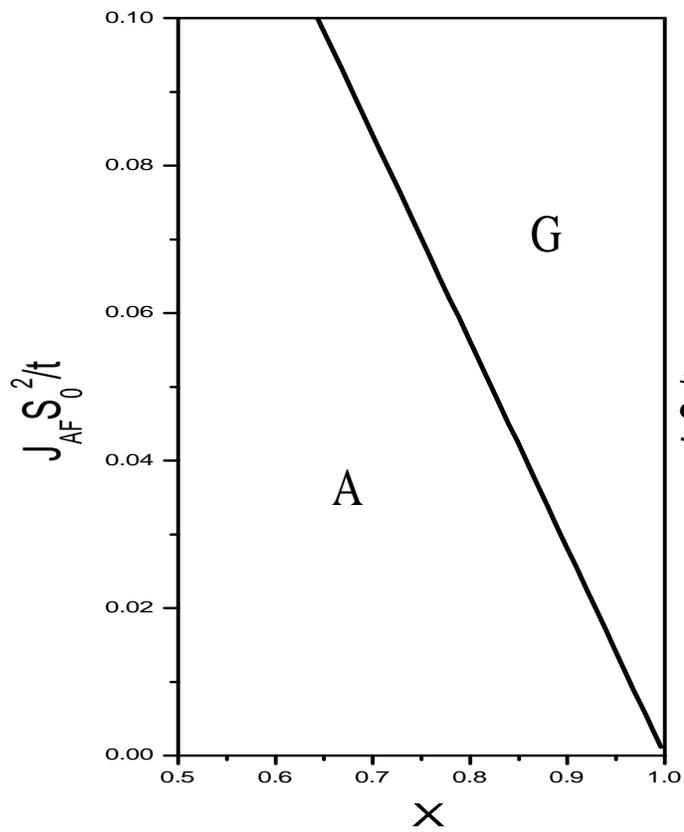
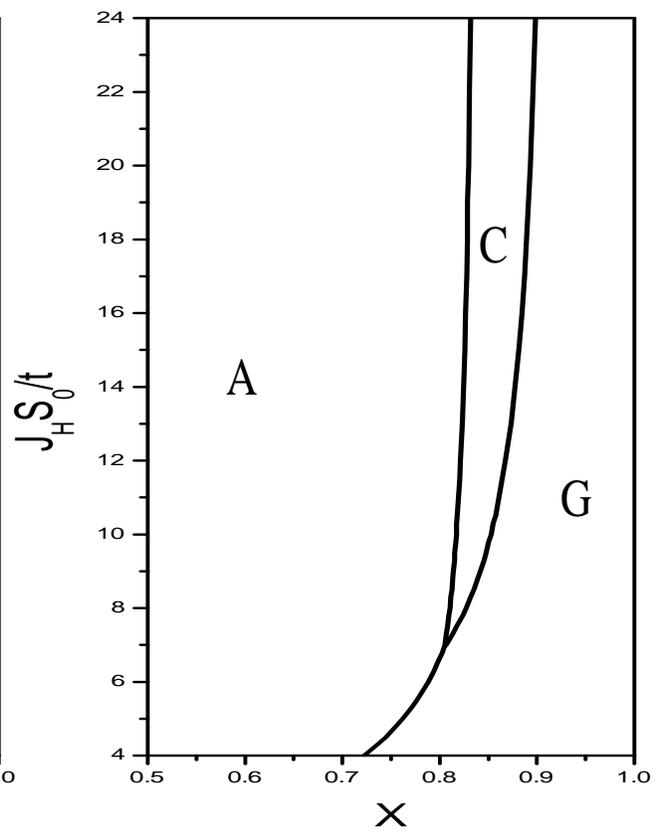
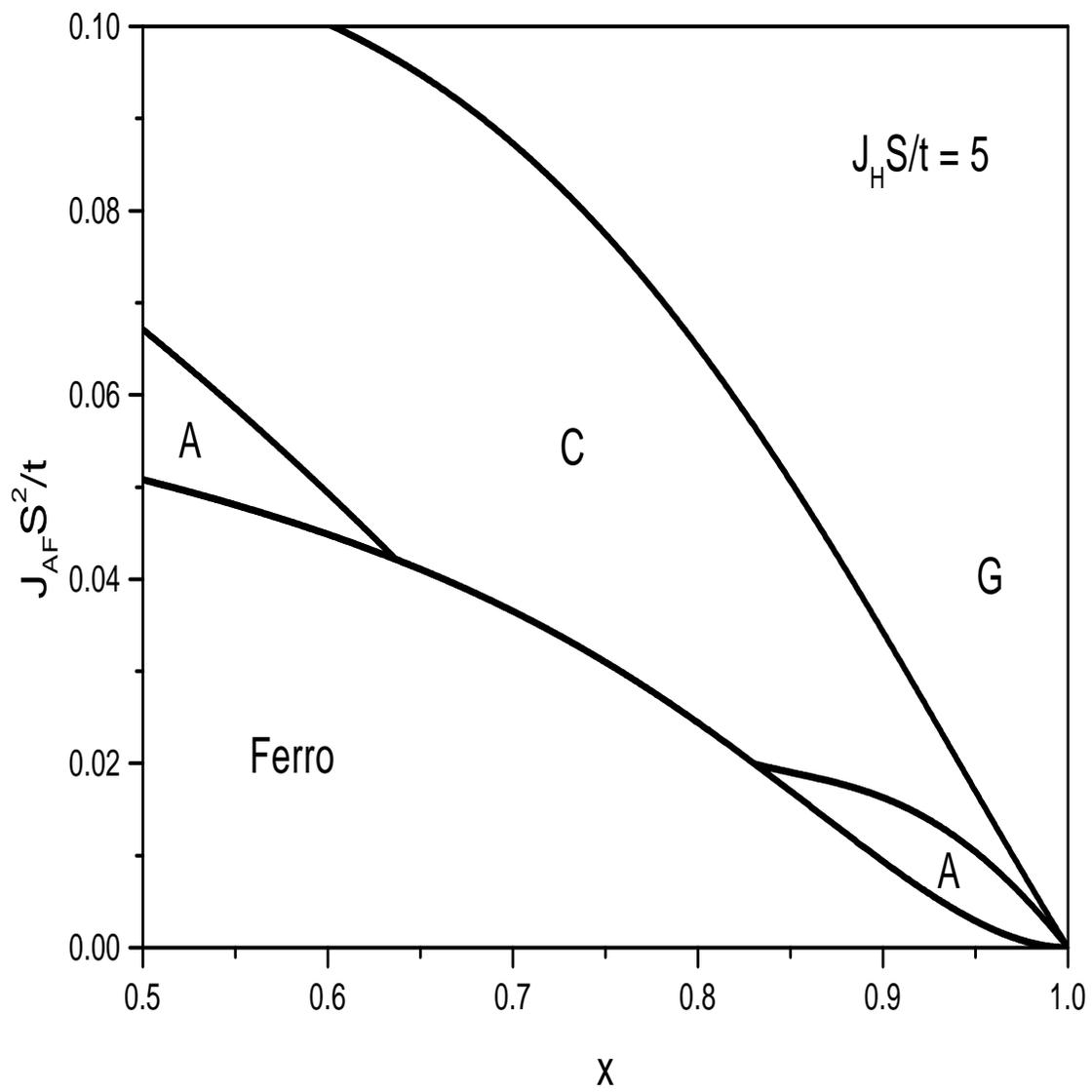
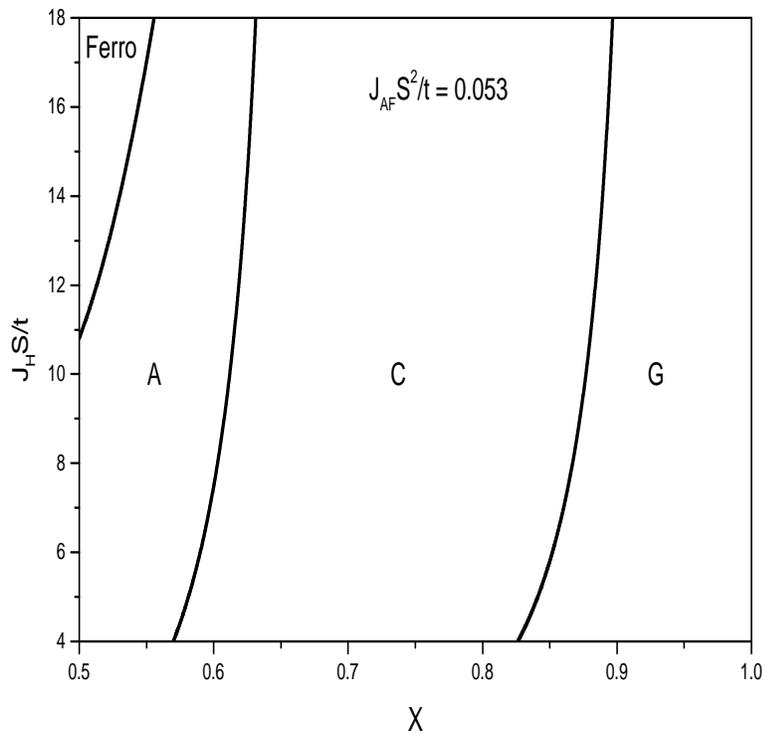
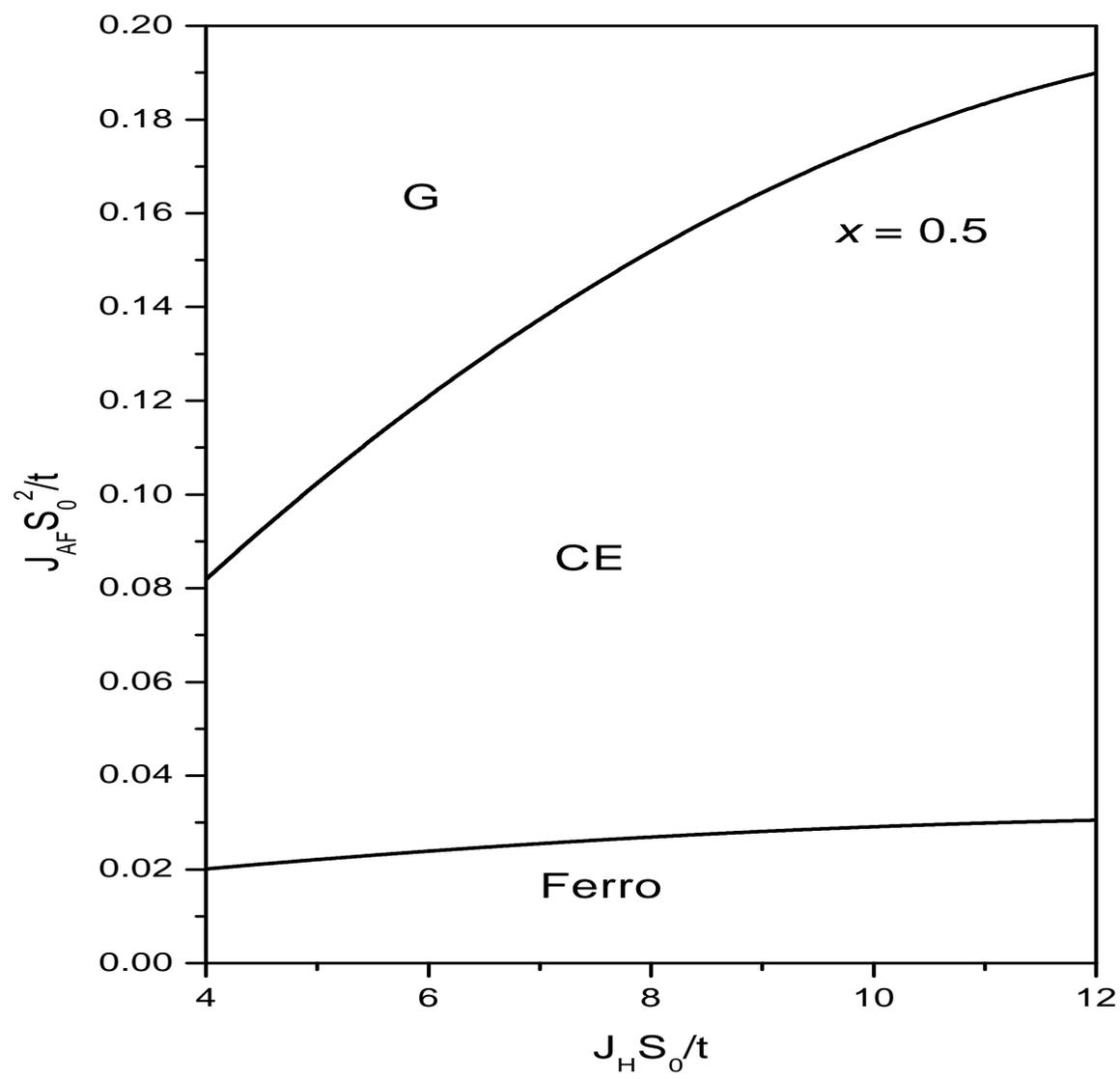
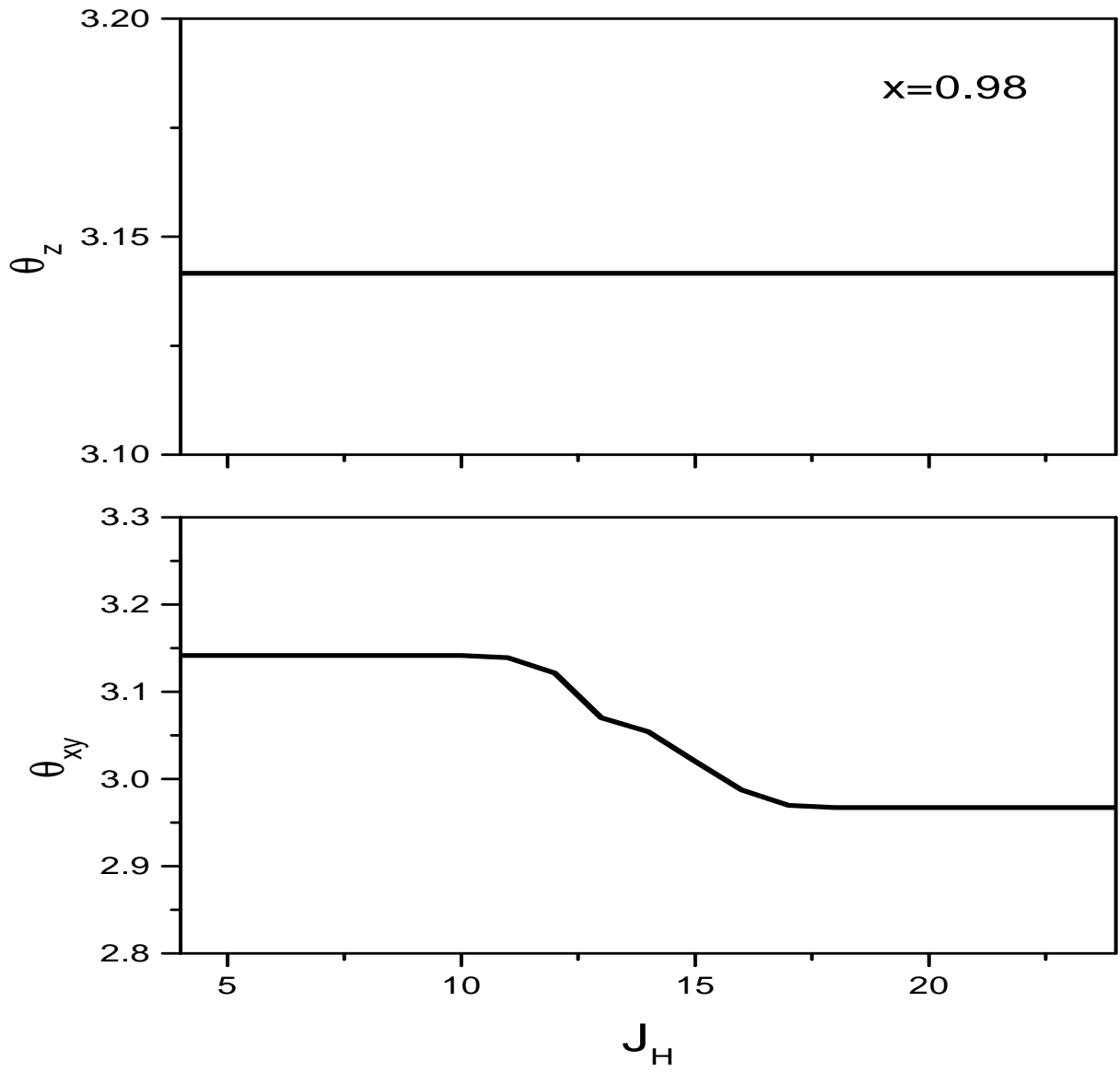
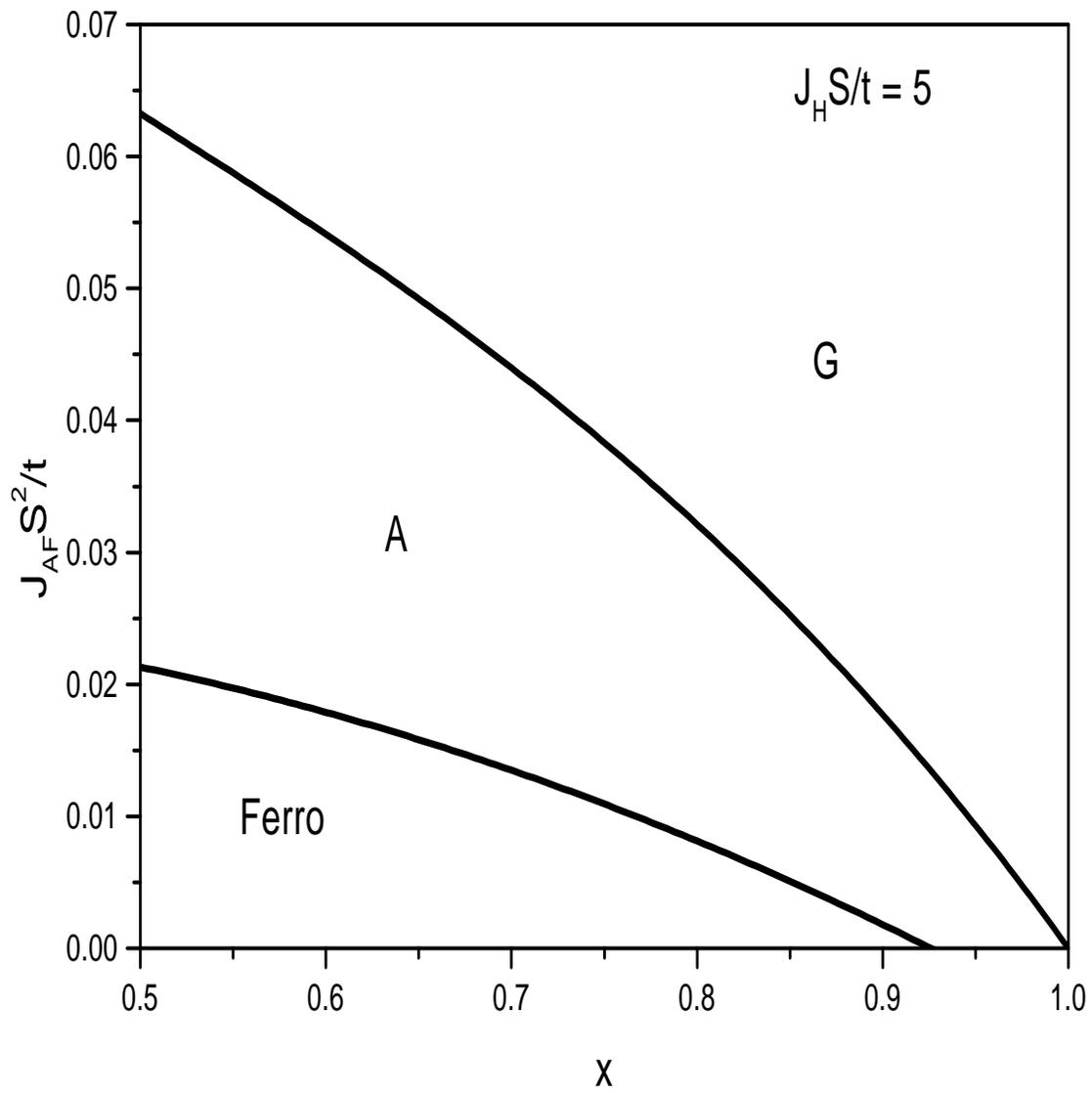
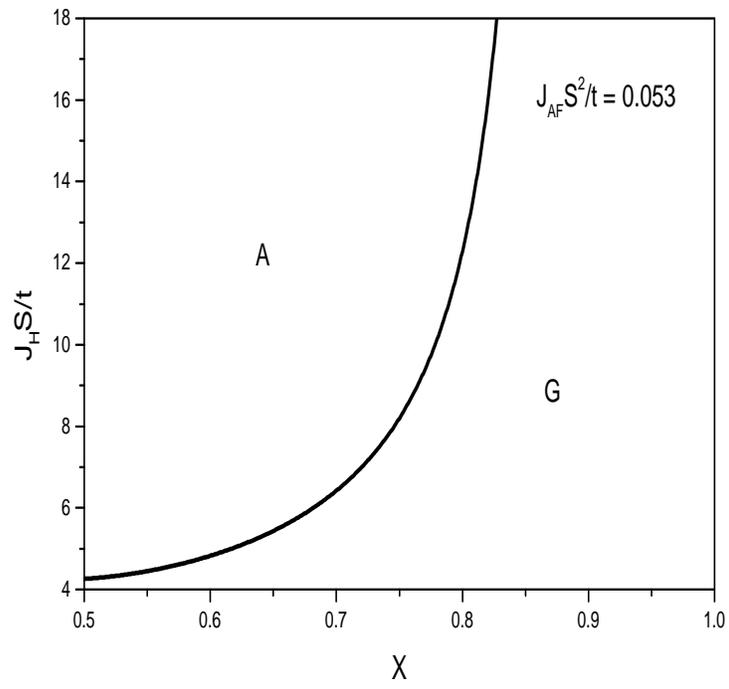
\end{document}